\documentclass[twocolumn,english,prl,showpacs,preprintnumbers,superscriptaddress]{revtex4-2}
\usepackage[utf8]{inputenc}
\usepackage{color}
\usepackage{amsmath}
\usepackage{amssymb}
\usepackage{graphicx}
\usepackage{textcomp}% Include figure files
\usepackage{dcolumn}% Align table columns on decimal point
\usepackage{bm}% bold math
\usepackage[right]{eurosym}
\usepackage{float}
\usepackage[english]{babel}
\usepackage{blindtext}
\usepackage{babel}
\usepackage[normalem]{ulem}
\usepackage{mhchem}
\usepackage{CJK}

\newcommand{\funit}{ph/\ensuremath{\mu}m\ensuremath{^2}}

\newcommand{\fluence}[2]{#1\ensuremath{\times}10\ensuremath{^{#2}}~\funit}
\newcommand{\ion}[2]{#1\ensuremath{^{#2+}}}
\begin{document}
\begin{CJK*}{UTF8}{}

%TC:ignore --- TeXcount: ignore title, author list, and abstract 

\title{Resonance-enhanced multiphoton ionization in the x-ray regime}

% What about full names here? Also I would like to put my Korean name here.
\author{Aaron C. LaForge}
\email{aaron.laforge@uconn.edu}
\affiliation{Department of Physics, University of Connecticut, Storrs, Connecticut 06269, USA}
\author{Sang-Kil Son \CJKfamily{mj}(손상길)}
\email{sangkil.son@cfel.de}
\affiliation{Center for Free-Electron Laser Science CFEL, Deutsches Elektronen-Synchrotron DESY, 22607 Hamburg, Germany}
\affiliation{The Hamburg Centre for Ultrafast Imaging, 22761 Hamburg, Germany}
\author{Debadarshini Mishra}
\affiliation{Department of Physics, University of Connecticut, Storrs, Connecticut 06269, USA}
\author{Markus Ilchen}
\affiliation{European XFEL, 22869 Schenefeld, Germany}
\affiliation{Institut f{\"u}r Physik und CINSaT, Universit{\"a}t Kassel, 34132 Kassel, Germany}
\author{Stephen Duncanson}
\affiliation{Department of Physics, University of Connecticut, Storrs, Connecticut 06269, USA}
\author{Eemeli Eronen}
\affiliation{Department of Physics and Astronomy, University of Turku, 20014 Turku, Finland}
\author{Edwin Kukk}
\affiliation{Department of Physics and Astronomy, University of Turku, 20014 Turku, Finland}
\author{Stanislaw Wirok-Stoletow}
\affiliation{Center for Free-Electron Laser Science CFEL, Deutsches Elektronen-Synchrotron DESY, 22607 Hamburg, Germany}
\affiliation{Department of Physics, Universit{\"a}t Hamburg, 22607 Hamburg, Germany}
\author{Daria Kolbasova}
\affiliation{Center for Free-Electron Laser Science CFEL, Deutsches Elektronen-Synchrotron DESY, 22607 Hamburg, Germany}
\affiliation{Department of Physics, Universit{\"a}t Hamburg, 22607 Hamburg, Germany}
\author{Peter Walter}
\affiliation{Linac Coherent Light Source, SLAC National Accelerator Laboratory, Menlo Park, California 94025, USA}
\author{Rebecca Boll}
\affiliation{European XFEL, 22869 Schenefeld, Germany}
\author{Alberto De~Fanis}
\affiliation{European XFEL, 22869 Schenefeld, Germany}
\author{Michael Meyer}
\affiliation{European XFEL, 22869 Schenefeld, Germany}
\author{Yevheniy Ovcharenko}
\affiliation{European XFEL, 22869 Schenefeld, Germany}
\author{Daniel E.\ Rivas}
\affiliation{European XFEL, 22869 Schenefeld, Germany}
\author{Philipp Schmidt}
\affiliation{European XFEL, 22869 Schenefeld, Germany}
\author{Sergey Usenko}
\affiliation{European XFEL, 22869 Schenefeld, Germany}
\author{Robin Santra}
\affiliation{Center for Free-Electron Laser Science CFEL, Deutsches Elektronen-Synchrotron DESY, 22607 Hamburg, Germany}
\affiliation{The Hamburg Centre for Ultrafast Imaging, 22761 Hamburg, Germany}
\affiliation{Department of Physics, Universit{\"a}t Hamburg, 22607 Hamburg, Germany}
\author{Nora Berrah}
\affiliation{Department of Physics, University of Connecticut, Storrs, Connecticut 06269, USA}

\begin{abstract}
Here, we report on the nonlinear ionization of argon atoms in the short wavelength regime using ultraintense x rays from the European XFEL. After sequential multiphoton ionization, high charge states are obtained. For photon energies that are insufficient to directly ionize a $1s$ electron, a different mechanism is required to obtain ionization to \ion{Ar}{17}. We propose this occurs through a two-color process where the second harmonic of the FEL pulse resonantly excites the system via a $1s \rightarrow 2p$ transition followed by ionization by the fundamental FEL pulse, which is a type of x-ray resonance-enhanced multiphoton ionization (REMPI). This resonant phenomenon occurs not only for \ion{Ar}{16}, but also through lower charge states, where multiple ionization competes with decay lifetimes, making x-ray REMPI distinctive from conventional REMPI. With the aid of state-of-the-art theoretical calculations, we explain the effects of x-ray REMPI on the relevant ion yields and spectral profile.
\end{abstract}

\date{\today}

\maketitle

%TC:endignore --- TeXcount

\end{CJK*}

%\begin{figure}
%	\begin{center}{
%			\includegraphics[width=0.5\textwidth]{Figures/Fig1.pdf}}
%		\caption{Electron kinetic energy distributions of resonantly excited He droplets for XUV-UV pump-probe delays of i) -1.4 ps, ii) 0 ps, and iii) 50 ps. The droplet size was 100,000 atoms and the photon energy was 21.6 eV.}
%		\label{fig1}
%	\end{center}
%\end{figure}

X-ray free-electron lasers (XFELs)~\cite{Pellegrini2016,Seddon2017} have offered new avenues for studying matter under intense femtosecond radiation. %where numerous photons can be absorbed within a few femtoseconds. 
In this case, the system is highly excited or ionized during the XFEL pulse resulting in a multitude of new processes such as nuclear resonance superradiance\,\cite{Chumakov2018}, plasma dynamics in solids\,\cite{Kluge2018}, and structural dynamics in complex molecular systems\,\cite{Nass2020}. 
Simpler systems such as atoms\,\cite{Young2010,Doumy2011,Rudek2012,Fukuzawa2013,Rudek2018}, molecules\,\cite{Hoener2010,Cryan2010,Rudenko2017}, and clusters\,\cite{Thomas2012,Gorkhover2012} provide a means to investigate multiphoton processes in a well-controlled manner.
%On the other side of the spectrum, 
For longer wavelengths, multiphoton processes have been well-studied. In particular, resonance-enhanced multiphoton ionization (REMPI)~\cite{Lambropoulos1976,Johnson1981,Miller1984} has been a useful spectroscopic technique, since it provides a highly sensitive and selective means to ionize molecular compounds without strong fragmentation effects.
As such, REMPI has been used in a wide variety of physical, chemical, and biological systems with recent applications in ultracold atom-molecule\,\cite{Akerman2017} and atom-atom\,\cite{Wolf2017} collisions, molecular chirality\,\cite{Vitanov2019}, catalytic surface chemistry\,\cite{Shirhatti2018}, nuclear-spin conversion\,\cite{Sugimoto2011}, and metrology\,\cite{Germann2014,Sinhal2020}. 

REMPI relies on the laser wavelength being tuned to an electronic resonance, which leads to promotion to an excited state followed by ionization. 
%REMPI can either be used with a single wavelength (one-color) or two wavelengths (two-color). 
As the absorption cross section for resonant excitation is typically stronger than the ionization cross section, REMPI provides an efficient means to ionize these systems at photon energies below the ionization threshold. 
Although XUV FELs have been used to show REMPI for shorter wavelengths~\cite{Shigemasa2013,Nikolopoulos2015}, extending REMPI to the x-ray regime requires entirely different physical processes and interpretation. Conventional REMPI relies on the excitation of a valence electron where the only relaxation pathway is radiative decay. On the other hand, a core-excited electron after x-ray resonant excitation can additionally relax by Auger-Meitner~\cite{Matsakis2019} decay, which is orders of magnitude faster than radiative decay. Thus, the complex interplay between ultrafast decay processes and REMPI renders this fundamental nonlinear process challenging to fully resolve in the x-ray regime.

\begin{figure}
	\begin{center}{
			\includegraphics[width=\linewidth]{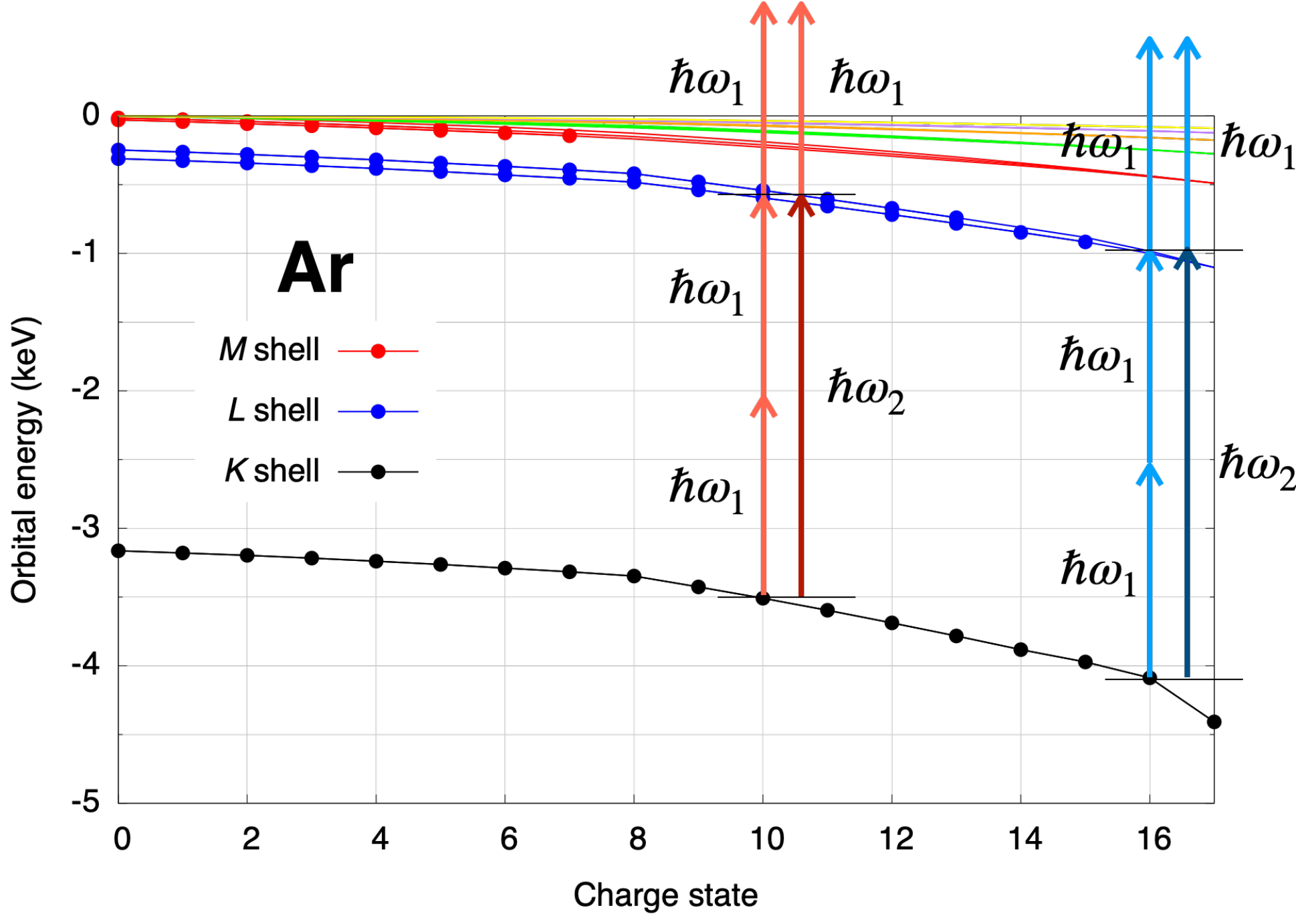}}
		\caption{(color online) Schematic of resonance-enhanced multiphoton ionization processes.
		The dark-colored arrows correspond to the second harmonic of the light-colored arrows, i.e., $\hbar\omega_2 = 2 \hbar\omega_1$. 
		The photon energies exemplified by the red arrows are smaller than the blue arrows.
		The orbital energies of the different electron shells are plotted as a function of the Ar charge state. See text for additional details.
		%The dark and light blue arrows signify the two resonant ionization paths for \ion{Ar}{16}.
		}
		\label{fig:schematic}
	\end{center}
\end{figure}

\begin{figure}
	\begin{center}{
			\includegraphics[width=1.0\linewidth]{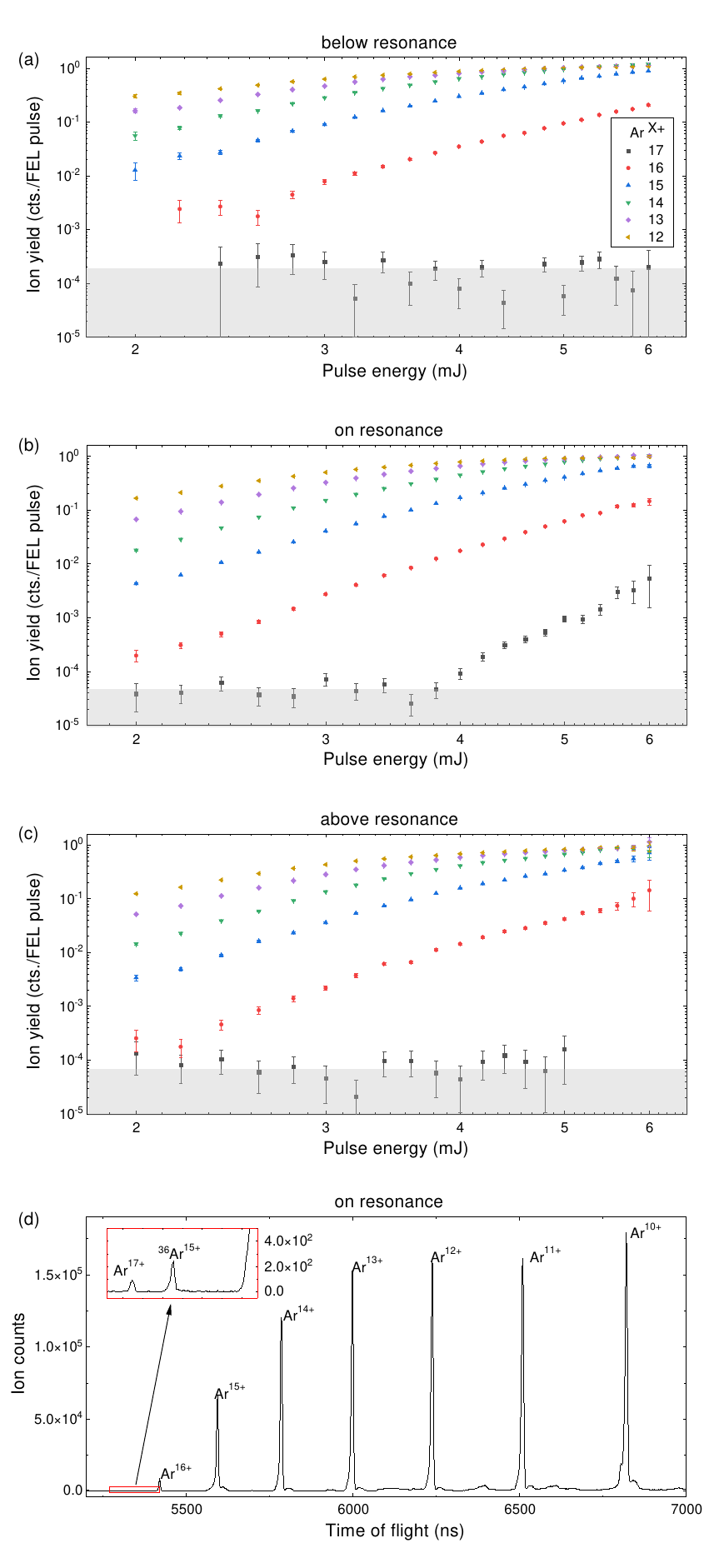}
		\caption{(color online) Ion yields of Ar as a function of the measured FEL pulse energy for three cases: (a) below the resonance, $\hbar\omega_1\,=\,1450$\,eV; (b) on the resonance, $\hbar\omega_1\,=\,1550$\,eV; and (c) above the resonance, $\hbar\omega_1\,=\,1576$\,eV. The ion yields are given in counts per FEL pulse. The experimental noise for each measurement is given as a gray area. (d) Ion time-of-flight spectrum for $\hbar\omega_1\,=\,1550$\,eV integrated over a measured pulse energy range of 4\,mJ--6\,mJ with the inset centered on \ion{Ar}{17} and an isotope of \ion{Ar}{15}.}
		\label{fig:ion_yields}}
	\end{center}
\end{figure}

In this Letter, we observe nonlinear ionization to create \ion{Ar}{17}, through a resonant two-color process in the x-ray regime. 
With the aid of state-of-the-art theoretical modeling, we attribute the ionization to a two-color REMPI-like process where the second harmonic of the FEL~\cite{Geloni2007} creates a $1s \rightarrow 2p$ core excitation and the fundamental FEL pulse subsequently ionizes the system. 
The core excitation can occur for charge states up to \ion{Ar}{16} where the lower charge states are influenced by their respective core-hole lifetimes.
We find the observed broadband nature of the spectral resonance to be due to overlapping resonances with lower Ar charge states.
%\sout{Conceptually, x-ray REMPI relies on an interplay between the steps of the ionization process and the lifetimes of Auger decay. As core electrons are excited, they can either be ionized by REMPI or relax by Auger decay on timescales of a few femtoseconds.}
%\textcolor{red}{\sout{X-ray REMPI, or XREMPI, unlike conventional REMPI, has the powerful attribute of element specificity in molecules, allowing selection of specific atoms in a molecule, thereby opening up a new spectroscopic technique to study x-ray ionization processes.}}

A schematic of the x-ray REMPI or XREMPI process is given in Fig.~\ref{fig:schematic}. %\textcolor{red}{\sout{where the orbital energies of the electron shells are plotted as a function of the Ar charge states.}}
For the higher charge states of Ar, the orbital energies shift to more negative values (higher binding energies).
The relative shifts in energies between shells lead to new resonances occurring, dependent on the charge state and the photon energy. 
%\textcolor{red}{\sout{In general, the resonance energy between the $K$ and $L$ shells shifts to higher energies for the higher charge states.}}
To illustrate this effect, we show the transitions for \ion{Ar}{10} and \ion{Ar}{16} with red and blue arrows, respectively, where the resonance shifts from $\sim$3000~eV to $\sim$3100~eV. 
To examine the formation of \ion{Ar}{17}, let us first consider the transition \ion{Ar}{16}$\rightarrow$\ion{Ar}{17}, which has an ionization energy of 4130 eV\,\cite{Artemyev2005}. 
For the case of an FEL photon energy of $\hbar\omega_1\,=\,1550$\,eV, \ion{Ar}{16} is produced through sequential multiphoton ionization~\cite{Santra2016}, but \ion{Ar}{17} cannot be produced through the absorption of another $\hbar\omega_1$ photon.
Instead, resonant excitation can be achieved through either a direct $\hbar\omega_1$ two-photon process (excited to $1s2s$) or an $\hbar\omega_2$ one-photon process (excited to $1s2p$), where $\hbar\omega_2$ is the second harmonic of the FEL pulse ($\hbar\omega_2 = 2 \hbar\omega_1$). 
These two processes are respectively represented by light and dark blue arrows in Fig.~\ref{fig:schematic}. 
Then, the resonantly excited states can be ionized by absorbing another $\hbar\omega_1$ photon.
As such, the overall pathway occurs through either a one-color, three-photon (2+1)-REMPI or a two-color, two-photon (1$'$+1)-REMPI, where the prime indicates the second harmonic.

Our experiment was performed using the Small Quantum Systems scientific instrument at the European XFEL~\cite{Tschentscher2017,Mazza2020}. 
In brief, soft x-ray FEL pulses with a nominal pulse length of 25\,fs were focused to a size of approximately 1.5\,$\mu$m$\times$1.5\,$\mu$m (FWHM) in the interaction region. 
The photon energy of the fundamental pulse was scanned from 1450~eV to 1583~eV with a measured energy bandwidth of approximately 1\,\% (FWHM).
The pulse energy varied from 2\,mJ to 6\,mJ due to shot-by-shot fluctuations within the pulse train, which was measured upstream by an x-ray gas monitor detector~\cite{Maltezopoulos2019}.
From a previous measurement, the contribution of second harmonic radiation was estimated to be between 0.2 and 0.6\,\%~\cite{Tanaka2021}. 
%That said, the contribution of the second harmonic can strongly vary between experiments, depending on the FEL beam parameters and electron energy.
Due to transmission of the grazing-incidence mirrors, third harmonic radiation is fully suppressed ($4\cdot10^{-9}$) for the photon energies used in this experiment~\cite{Mazza2021}.
The FEL beam crossed an effusive beam of Ar gas in the interaction region where an ion time-of-flight spectrometer was used to count ions on a shot-by-shot basis. 
%By modeling with \textsc{xcalib}~\cite{Toyota2019}, the peak fluence at 1550~eV was calibrated to be \fluence{1.6}{12} and \fluence{2.3}{12} by measuring the ion yields of Ne and Ar (charge states less than $+16$) as reference gases, respectively, assuming a single Gaussian spatial fluence distribution in the interaction region.
%If we assume a focal area of 1.5\,$\mu$m$\times$1.5\,$\mu$m, the corresponding beamline transmissions are 17.1\% (Ne) and 23.8\% (Ar).
%Since the calibration determines only the ratio between the transmission, measured to be about 70\% at these photon energies, and the focal area, the simulations would indicate a somewhat larger focal area or a non-Gaussian beam profile.
To analyze and interpret our data, we performed theoretical calculations using \textsc{xatom}~\cite{Son2011,Jurek2016,Toyota2019}, which solves coupled rate equations to simulate x-ray multiphoton multiple ionization dynamics with decay rates and cross sections calculated using the Hartree--Fock--Slater approach.
To calculate direct two-photon and one-photon resonant absorption cross sections, we employed grid-based time-dependent configuration interaction singles~\cite{Sytcheva2012,Pabst2017} and an extended version of \textsc{xatom}~\cite{Toyota2017}, respectively.

Figure~\ref{fig:ion_yields} shows the ion yields of the high charge states ($>+10$) of Ar as a function of the measured pulse energy for three cases: (a) below the resonance, (b) on the resonance, and (c) above the resonance.
For completeness, the ion time-of-flight spectrum on the resonance is shown in Fig.~\ref{fig:ion_yields}\,(d) integrated over a measured pulse energy range of 4\,mJ--6\,mJ.
The figure shows the higher charge states of Ar with \ion{Ar}{17} and $^{36}$\ion{Ar}{15} shown in the inset.
%Here, the resonance indicates a transition from the $K$ shell to the $L$ shell, as depicted in Fig.~\ref{fig:schematic}.
For Figs.~\ref{fig:ion_yields}(a)--(c), the data are plotted on a log-log scale such that the slope of the distribution is proportional to the number of absorbed photons, as long as there is no saturation~\cite{Makris2009,Richter2010,Santra2016}.
%As the pulse energy increases, the ion yields first increase, and the yields of lower charge states become saturated when they are depleted.%~\cite{Makris2009,Richter2010,Santra2016}.
%Since the interaction volume has a fluence distribution with sufficiently broad low-fluence areas, the slope asymptotically approaches zero, rather than reaching negative values.%~\cite{Nikolopoulos2013,Toyota2019}.
%This behavior has the result that when the state \ion{Ar}{q} begins to delineate, the higher states \ion{Ar}{(q+1)} additionally saturate at slightly higher pulse energies.
Overall, the ion yields show very similar power dependences for all three cases, with the exception of \ion{Ar}{17}. 
%\textcolor{red}{\sout{As can be seen in Fig.~\ref{fig:ion_yields},}} 
For both above and below the resonance, the \ion{Ar}{17} yield shows no dependence on the pulse energy. 
Given the statistical error, the yields fall within the experimental noise (gray area). %\textcolor{red}{\sout{, which are shown as gray areas in Fig.~\ref{fig:ion_yields} for each measurement}}. 
On the other hand, when the photon energy is tuned to the resonance, the \ion{Ar}{17} yield shows an increase for pulse energies greater than 4\,mJ. 

\begin{figure}
	\begin{center}{
			\includegraphics[width=\linewidth]{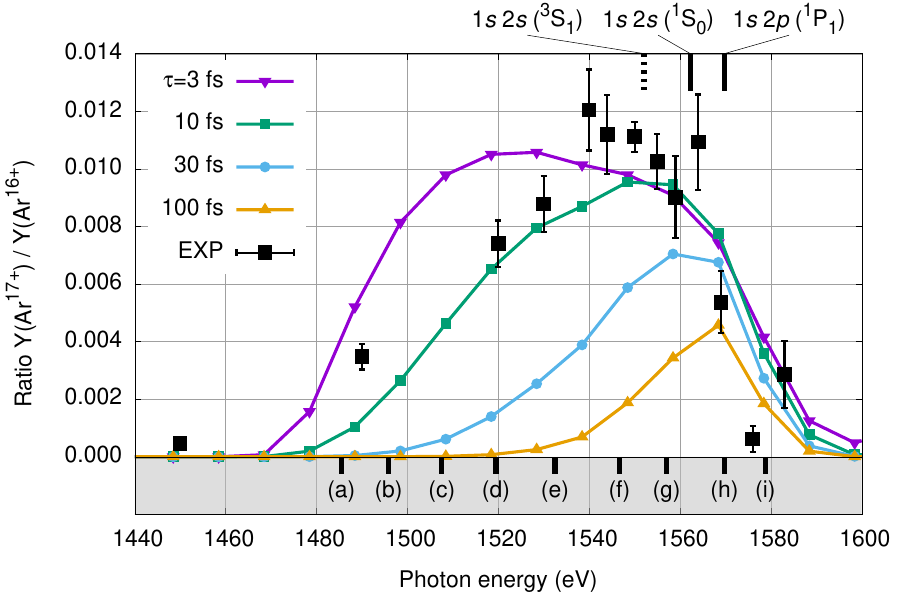}}
		\caption{(color online) \ion{Ar}{17} to \ion{Ar}{16} yield ratio as a function of the photon energy at a pulse energy of 4.2$\pm$0.1\,mJ. Theoretical yield ratios for different pulse lengths with a fixed pulse energy of 4.2\,mJ are given in different colors. 
		The labels at the bottom indicate $1s \rightarrow 2p$ transitions for different charge states and electronic configurations. %corresponding to the peaks for the 10-fs pulse length case in Fig.~S7(a). %The label of ($l$,$m$) refers to the configuration of $K^2 L^l M^m$, whose charge is $q=16-l-m$.
%		a) Ar$^{16+}$: $1s^2 \rightarrow 1s 2p$, 
%		b) Ar$^{15+}$: $1s^2 3p \rightarrow 1s 2p 3p$, 
%		c) Ar$^{15+}$: $1s^2 2p \rightarrow 1s 2p^2$, 
%		d) Ar$^{14+}$: $1s^2 2p^2 \rightarrow 1s 2p^3$, 
%		e) Ar$^{11+}$: $1s^2 2s 2p^2 3s^2 \rightarrow 1s 2s 2p^3 3s^2$, 
%		f) Ar$^{10+}$: $1s^2 2s 2p^3 3s^2 \rightarrow 1s 2s 2p^4 3s^2$, 
%		g) Ar$^{9+}$: $1s^2 2s 2p^4 3s^2 \rightarrow 1s 2s 2p^5 3s^2$, 
%		h) Ar$^{6+}$: $1s^2 2s^2 2p^4 3p^4 \rightarrow 1s 2s^2 2p^5 3p^4$, and
%		i) Ar$^{4+}$: $1s^2 2s^2 2p^5 3p^5\rightarrow 1s 2s^2 2p^6 3p^5$.
		(a) Ar$^{4+}$: $1s^2 2s^2 2p^5 3p^5$.
		(b) Ar$^{6+}$: $1s^2 2s^2 2p^4 3p^4$,
		(c) Ar$^{9+}$: $1s^2 2s 2p^4 3s^2$, 
		(d) Ar$^{10+}$: $1s^2 2s 2p^3 3s^2$, 
		(e) Ar$^{11+}$: $1s^2 2s 2p^2 3s^2$, 
		(f) Ar$^{14+}$: $1s^2 2p^2$, 
		(g) Ar$^{15+}$: $1s^2 2p$, 
		(h) Ar$^{16+}$: $1s^2$, and
		(i) Ar$^{15+}$: $1s^2 3p$.
		Here, the electron configurations are given before the transition.
		See text for additional details.}
		\label{fig:resonance_profile}
	\end{center}
\end{figure}
 
To further investigate this ionization process, we have measured the resonance spectral profile. 
Figure~\ref{fig:resonance_profile} presents the experimental yield ratio of \ion{Ar}{17} to \ion{Ar}{16} as a function of the photon energy at a pulse energy of 4.2$\pm$0.1\,mJ.
The expected resonance positions, corresponding to half of the $1s2s\;(^1\text{S}_0)$ and $1s2p\;(^1\text{P}_1)$ transition energies of \ion{Ar}{16}~\cite{Saloman2010}, are marked as solid vertical lines on the top of the figure.
The $1s2s\;(^3\text{S}_1)$ state (dashed line) does not have a significant contribution to the resonance since the two-photon transition to this state is forbidden and the M1 one-photon transition is suppressed~\cite{Savukov2003}.
%At the peak intensity under consideration, the AC Stark shift on the resonance is negligible [see Fig.~S1 in the Supplemental Material (SM) for additional details].
Nonetheless, our experimental data clearly show that a) the peak position is red-shifted compared to the literature values, and b) the distribution is broad and asymmetric toward lower photon energies. 
%Note that the observed red-shift is outside of the experimental error bounds.
%Note that the measured photon energy is within a few eV of the specified value, so the observed red-shift is outside of the experimental error bounds.
These features are in stark contrast to the typical spectral dependence of REMPI at longer wavelengths, which gives sharp, narrow peaks when the laser is on the resonant transition. 
What causes these distinctions for XREMPI compared to conventional REMPI?

\begin{figure}
	\begin{center}{
			\includegraphics[width=\linewidth]{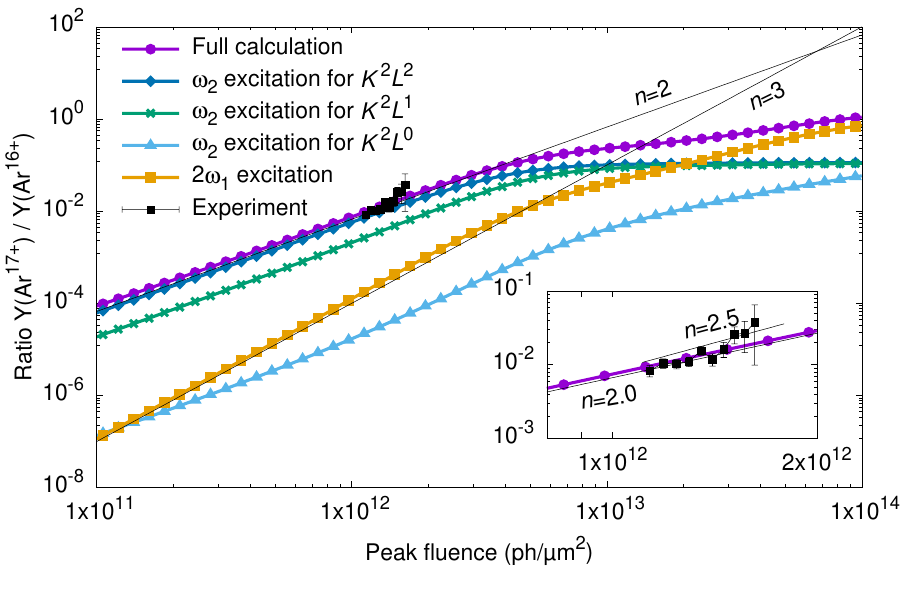}}
		\caption{(color online) \ion{Ar}{17} to \ion{Ar}{16} yield ratio as a function of the peak fluence at a photon energy of 1550~eV. Theoretical yield ratios are calculated including various contributions. See text for additional details. For the experimental data (black), the pulse energy is converted into the peak fluence based on a calibration using the ion yields of Ne~\cite{Toyota2019}.
		Inset: Zoomed near the peak fluence of \fluence{1.6}{12} with fitted slopes.
%		{\color{blue}[This figure is generated with 0.1\% of 2nd harmonic contribution, but I think we'd better put 0.2\%. Then, we may use the Ne calibrated peak fluence.]}
		}
		\label{fig:slope}
	\end{center}
\end{figure}

We first examine the pulse-length dependence of the resonance profile by using theoretical modeling in Fig.~\ref{fig:resonance_profile}. 
The yield ratios are calculated for several pulse lengths, which are given as multiple colors.
The dependence on other parameters is given in Figs.~S1--S3 in the Supplemental Material (SM). 
Besides one-photon ionization by the fundamental pulse, one-photon ionization and resonant excitation by the second harmonic are included in the calculations for all charge states.
We assume that the second harmonic contribution is 0.2\% of the fundamental fluence and its bandwidth is equivalent to the fundamental bandwidth, which was measured to be $\sim$\,1\% for this experiment.
On the other hand, direct two-photon excitation by the fundamental pulse is excluded since its contribution turns out to be negligible at fluences under consideration (see Fig.~\ref{fig:slope}).
The theoretical results are shifted on the $x$ axis by the difference between one half of the $1s2p$ transition energy of \ion{Ar}{16} calculated using \textsc{xatom} (1551.3~eV) and the literature value (1569.8~eV)~\cite{Saloman2010}.
The results are volume-integrated~\cite{Toyota2019} with a pulse energy fixed at 4.2\,mJ.
Our calculations demonstrate that the peak position is shifted to lower photon energies and the profile width is broadened as the pulse length becomes shorter.
Note that these effects on the resonance profile are not strongly sensitive to the calibrated peak fluence or transmission (see Fig.~S1 in the SM for their resonance profile dependence). 
The calculated resonance profile with a pulse length of 10~fs fairly matches the experimental profile, which is consistent with a prior observation that the x-ray pulse length is a few times shorter than its nominal value~\cite{Young2010}.

%\textcolor{cyan}{For conventional REMPI, the spectral profile is determined by the lifetime of the resonant state. Broadening due to the bandwidth of the laser is symmetric and related to the pulse length by the Fourier transform limitation. For x-ray REMPI, the broadening we observe is asymmetric and redshifted, which is essentially independent of the FEL bandwidth (see Fig.~S4 in the SM). The observed feature in the spectral profile is directly due to the core excitations from the lower charge states. In other words, there is a competition between the ionization steps and Auger decay. In that regard, the FEL pulse length plays a critical role in determining the spectral profile (see Figs.~\ref{fig:resonance_profile} and S7 in the SM).} Only when the pulse length becomes sufficiently long ($\sim$\,100\,fs) does the resonance profile peak at the resonance corresponding to the $1s2p$ transition of \ion{Ar}{16} and its width is reduced to the FEL bandwidth of $\sim$\,30~eV \textcolor{red}{($\sim$\,1\% of the second harmonic)}, thereby resembling a conventional REMPI spectral width.
%\textcolor{red}{Thus, the conventional REMPI picture is \textcolor{blue}{applicable only} for the long-pulse-length limit in the x-ray regime.} 

For a conventional REMPI experiment using ultrashort laser pulses, the resonance profile is governed by the homogeneous bandwidth broadening due to the pulse length being shorter than the lifetime of the resonant state.
In this case, the resonance profile can also be shifted due to the AC Stark effect since the intensity increases as the pulse length decreases, but the resonance peak still remains symmetric.
%\sout{Even though the anticipated pulse length is shorter than the lifetime of the \ion{Ar}{16} $1s2p$ state ($\sim$\,35~fs), the measured resonance profile and calculated pulse-length dependence in Fig.~\ref{fig:resonance_profile} cannot be explained by the conventional REMPI for the following reasons.}
Having said that, the measured resonance profile and its pulse-length dependence in Fig.~\ref{fig:resonance_profile} cannot be explained by conventional REMPI for the following reasons.
First, the observed width ($\sim$\,60~eV) is much broader than the FEL bandwidth, which corresponds to about 30~eV ($\sim$\,1\% of the second harmonic).
In fact, for a self-amplified spontaneous emission (SASE) XFEL pulse, the bandwidth is not directly related to the pulse length but determined by the shortest pulse duration of the ``spiky'' pulses~\cite{Kondratenko1980,Bonifacio1984,Andruszkow2000,Milton2001,Rohringer2007}. For our calculations, we employed the same FEL bandwidth for the various pulse lengths.
Second, the observed red-shift with respect to the resonances at \ion{Ar}{16} is more than 10~eV, but the AC Stark shift in the x-ray regime is negligible ($<$\,0.2~eV) at the peak intensity under consideration (see Fig.~S4 in the SM).
Third, the resonance profile is asymmetric towards lower photon energies, which strongly hints that resonances occur for charge states lower than \ion{Ar}{16}.
 When the pulse length becomes sufficiently long ($\sim$\,100\,fs), the resonance profile peaks at the resonance corresponding to the $1s2p$ transition of \ion{Ar}{16} and its width is reduced to the FEL bandwidth of $\sim$\,30~eV.

To get a better insight into the XREMPI mechanism, we analyze the contributions of individual resonant processes occurring for all charge states including \ion{Ar}{16}, at the peak of the resonance profile ($\hbar\omega\,=\,1550$\,eV) with a fixed pulse length of 10~fs.
Figure~\ref{fig:slope} shows the ion yield ratios of \ion{Ar}{17} to \ion{Ar}{16} as a function of the calibrated peak fluence.
Similar to Fig.~\ref{fig:ion_yields}, the experimental data are plotted on a log-log scale and fitted with a simple power function, $y=C \times x^n$, where $n$ is a measure of number of photons absorbed. 
The fit yields a power dependence of $n=2.5\pm0.6$, which points to a nonlinear ionization process by either three or two photons.
We additionally plot theoretical ion yield ratios calculated for the various XREMPI pathways in Fig.~\ref{fig:slope}.
The purple (circles) curve represents the full XREMPI calculation including one-photon resonant excitation by the second harmonic ($\omega_2$) and direct two-photon resonant excitation by the fundamental ($2\omega_1$), both of which are calculated for all charge states.
For the ground state of \ion{Ar}{16}, the direct two-photon cross section is explicitly calculated (see Fig.~S5 in the SM).
For the other charge states, we have assumed the same cross-section profiles, but shifted them according to the calculated transition energies.
The orange (squares) curve indicates the calculated yield assuming only the direct two-photon resonant excitation cross sections, which results in a power dependence of three.
The light blue (triangles) curve presents the calculation with the $\omega_2$ excitation of $1s \rightarrow 2p$ only for electron configurations $K^2 L^0 M^m$ where $0 \leq m \leq 8$.
The green (crosses) curve is for $K^2 L^1 M^m$, and the dark blue (diamonds) curve is for $K^2 L^2 M^m$.

Our analysis demonstrates that, even for the lower bound estimate of the second harmonic contribution (0.2\%), the one-photon resonant excitation of $K^2 L^2 M^m$ by the second harmonic predominantly contributes to the formation of \ion{Ar}{17} in the range of experimental peak fluences.
Note that $K^2 L^2 M^m$ corresponds to the charge state of $+q$, where $+6 \le q \le +14$.
In this case, a $1s$ vacancy is formed in the lower charge states, \ion{Ar}{q}, and must survive until the final ionization step, i.e., \ion{Ar}{17}.
Thus, its ionization mechanism is characterized by (1$'$+$n$)-XREMPI from \ion{Ar}{q} to \ion{Ar}{17}, where $n = 17 - q$.
This requires absorption of $(n+1)$ photons and ejection of $n$ electrons.
On the other hand, formation of ground state \ion{Ar}{16} from \ion{Ar}{q} requires $(16-q)=(n-1)$ photons.
Therefore, even for cases when $1s$ promotion occurs in the lower charge states, the \ion{Ar}{17} to \ion{Ar}{16} yield ratios plotted in Fig.~\ref{fig:slope} still yield a power dependence of two, except for the orange curve.

This (1$'$+$n$)-XREMPI process in the lower charge states can explain the asymmetric broadening of the resonance profile shown in Fig.~\ref{fig:resonance_profile} since the $1s \rightarrow 2p$ transition energy is shifted to smaller energies for the lower charge states, as shown in Fig.~\ref{fig:schematic}.
Furthermore, ultrafast decay processes are critical in x-ray multiphoton ionization dynamics~\cite{Santra2016}.
For example, the ion yields of Ar without decay processes are substantially reduced (see Fig.~S6 in the SM).
If the pulse length is comparable to or shorter than the decay lifetimes, it is more likely that the $1s$ vacancy formed in the lower charge states can be further ionized before the decay process occurs (see Fig.~S7 in the SM for Auger-Meitner and fluorescence lifetimes as a function of charge state).
Surprisingly, the most probable XREMPI transition for a pulse length of 10~fs occurs for Ar$^{14+}$. In that case, the second harmonic first excites the $1s$ electron, followed by the absorption of three photons from the fundamental pulse in order to reach Ar$^{17+}$, which would be (1$'$+3)-XREMPI.
This way, the early-formed vacancy can survive over the course of sequential ionization up to +17, thus leaving its footprint on the resonance profile of \ion{Ar}{17}/\ion{Ar}{16}.
This phenomenon resembles the quasinonsequential mechanism that is expected to lead to a breakdown of frustrated absorption~\cite{Ho2015,Son2020}.
Thus, the pulse-length sensitivity of the resonance profile is due to the various ionization pathways and associated decay lifetimes, rather than the bandwidth of the pulse.

It is worthwhile to compare XREMPI with resonance-enabled or enhanced x-ray multiple ionization (REXMI)~\cite{Rudek2012,Rudek2018}, one of the distinctive phenomena in the field of XFEL--matter interactions.
REXMI exploits a broad energy bandwidth to drive multiple resonant excitations for a range of charge states.
After multielectron excitations, further ionization occurs via electron-correlation-driven relaxation processes.
Due to the intrinsic broad bandwidth which characterizes SASE pulses~\cite{Kondratenko1980,Bonifacio1984,Andruszkow2000,Milton2001,Rohringer2007}, REXMI-type processes have been demonstrated as an ultra-efficient ionization mechanism in previous XFEL experiments.
On the other hand, XREMPI involves a single-electron excitation, and further ionization occurs only by absorbing additional photons.
Since XREMPI occurs only for a specific transition and charge state, it is beneficial to have a narrow bandwidth through the use of a monochromator or self-seeding techniques (see Fig.~S8 in the SM for the resonance profiles calculated from narrower bandwidths).

In conclusion, we observed a novel type of REMPI in the x-ray regime experimentally and validated the process theoretically.
Through sequential multiphoton ionization, neutral Ar is ionized to high charge states.
For photon energies that are insufficient to directly ionize $1s$ electrons, promotion to \ion{Ar}{17} requires a resonant process to create a $1s$ vacancy. 
We show that this occurs through a REMPI-like process where the second harmonic of the FEL promotes a $1s \rightarrow 2p$ transition and the fundamental FEL pulse subsequently ionizes the system further. 
That said, XREMPI is not restricted to a combination of the fundamental and second harmonic, and two-color capabilities at XFELs~\cite{Hara2013,Lutman2013,Serkez2020} could provide a desirable means to study XREMPI.
% For two-color XFELs
% Hara2013 (SACLA): Nat. Commun., first demonstration at SACLA
% Lutman2013 (LCLS): PRL, first demonstration at LCLS
% Lutman2016 (LCLS): Nat. Photon., slice, multicolor
% Lu2018 (EuXFEL): Rev. Sci. Instrum. development of facilities
% Serkez2020 (EuXFEL): Appl. Sci., review/current status for EuXFEL

The resonance spectral profile of the XREMPI process shows a broad, asymmetric, red-shifted distribution due to overlapping resonances with lower charge states, which is a clear distinction from conventional REMPI profiles at longer wavelengths.
We have also demonstrated the strong dependence of the resonance profile on the pulse length, which is potentially applicable to characterize FEL beam parameters.
With advances in seeding of XFELs and the availability of narrow bandwidth radiation~\cite{Amann2012,Hemsing2020}, XREMPI can be used to reveal individual resonance structures in the spectral profile. 
This capability can be applied to perform precision spectroscopy on atoms or molecules; for example, highly charged ions of astrophysical relevance~\cite{Bernitt2012,Kuhn2020}.
%\textcolor{cyan}{X-ray REMPI also offers the powerful new aspect of element specificity in molecules. For conventional REMPI, the spectral profiles show a strong wavelength dependence which can be enhanced by specific atomic sites in the molecules, thereby opening up a new spectroscopic technique to study x-ray ionization processes. Furthermore, with x-rays, one can selectively excite one atom in a molecule and probe another atom.}
XREMPI also offers the powerful new aspect of element specificity in molecules, thereby opening up a novel spectroscopic technique to study x-ray ionization processes.
For instance, with x rays, one can selectively excite one atom in a molecule and probe another atom.
One can take advantage of the XREMPI technique to determine many physical processes such as charge migration from one site in the molecule to another position, thereby tracking the response of several different atoms in the molecule.

Data recorded for the experiment at the European XFEL are available on request at \cite{EuXFEL_data}.

%TC:ignore --- TeXcount: ignore acknowledgment and references
\begin{acknowledgements}
A.C.L., D.M., S.D., and N.B.\ gratefully acknowledge financial support from the Chemical Sciences, Geosciences and Biosciences Division, Office of Basic Energy Sciences, Office of Science, U.S.\ Department of Energy, grant no.\ DE-SC0012376. 
P.W.\ acknowledges financial support from SLAC National Accelerator Laboratory, supported by the U.S.\ Department of Energy, Office of Science, Office of Basic Energy Sciences under contract no.\ DE-AC02-76SF00515.
E.K.\ acknowledges funding by the Academy of Finland.
M.I.\ and Ph.S.\ acknowledge funding by the Volkswagen foundation within a Peter-Paul-Ewald fellowship.
M.M.\ acknowledges funding by the Deutsche Forschungsgemeinschaft (DFG, German Research Foundation) - SFB-925 - project 170620586. 
The research leading to this result has been supported by the project CALIPSOplus under the Grant Agreement 730872 from the EU Framework Programme for Research and Innovation HORIZON 2020.
We acknowledge European XFEL in Schenefeld, Germany, for provision of x-ray free-electron laser beam time at the SQS instrument and would like to thank the staff for their assistance. 
We thank Prof.\ Alan Wuosmaa (UConn) for helpful discussions. 

A.C.L.\ and S.-K.S.\ contributed equally to this work.
\end{acknowledgements}	
	
%\bibliography{Ar_XFEL}
%apsrev4-2.bst 2019-01-14 (MD) hand-edited version of apsrev4-1.bst
%Control: key (0)
%Control: author (8) initials jnrlst
%Control: editor formatted (1) identically to author
%Control: production of article title (0) allowed
%Control: page (0) single
%Control: year (1) truncated
%Control: production of eprint (0) enabled
%

\end{document}